\documentclass[journal,onecolumn,10pt]{WSPLC_class}

\usepackage[left=2cm,top=1.78cm,right=2cm]{geometry}
\usepackage{graphicx}
\usepackage{verbatim}
\usepackage{amsmath}
\usepackage{subfig}
\begin{document}

\title{Broadband MIMO Couplers Characterization and Comparison}

\author{
	\large
	 Davide Righini\footnote{$^{(1)}$ The work of D. Righini was done at the University of Klagenfurt, Austria, and partially supported by the ERASMUS+ training program.}$^{(1)}$ and Andrea M. Tonello$^{(2)}$ \\\vspace{6pt}
	\normalsize
	 	$^{(1)}$Universit\`a di Udine - Italy, e-mail: davide.righini@gmail.com \\\vspace{4pt}
		$^{(2)}$Alpen-Adria-Universit{\"a}t Klagenfurt - EcoSys Lab - Austria, e-mail: andrea.tonello@aau.at \\\vspace{4pt}
}
% make the title area
\maketitle

\vspace{-10pt}

\IEEEpeerreviewmaketitle

\section{Introduction}
\IEEEPARstart{T}{his} paper focuses on MIMO Power Line Communication (PLC) to provide increased performance. In MIMO PLC, appropriate coupling methods are necessary in order to enable the effective injection of the signal through the broad band PLC channel so that high data rates can be achieved \cite{1,2,3}. In the literature, different kind of couplers are proposed  \cite{4},\cite{5}. In this paper, we want to characterize in an analytic way strengths and weaknesses of each coupler design (topology) from an electrical circuit perspective.
We dwell on the description and analysis of the three main and common configurations used for the MIMO couplers: star (S), triangle ($\Delta$) and T. For each configuration, we study different connection formats for the input generators and output loads: star, triangle and T. Furthermore, we consider MIMO 2x3, 2x4, 3x3, 3x4 architectures, where the fourth received signal is referred to as common mode signal. 

The purpose in this study is to clearly define the equations that describe the output signals as a function of the input signals and the impedances present in the circuit. This is achieved analyzing the transfer function of the circuit under survey.   
A model is presented to be able to identify how much the nonidealities of components, such as, uncoupling/protection devices and filters, affect the coupling and cross coupling between signals. All these components add, inside the transmission chain, undesired effects in the signal of interest.
Our aim is to define a procedure that, knowing the characteristics of these nonideal components, allows us to compensate this effects working on circuit parts and signals. 

Moreover, the effect of the line impedance variation on the coupled voltage is investigated. This is an important aspect since the line impedance may have a time variant behavior as a result of network loads that are time variant \cite{6}. In particular, we consider the impedance mismatch problem \cite{7} under a source load constraint with distinct coupler topologies: star, triangle and T. 

Finally, a comparison of the best combination of TX-RX coupler configurations is carried out. 

\section{MIMO Couplers Circuit analysis}
\begin{figure}[ht!]
	\centering 
	\subfloat[ \label{fig:star-star_coupler}]{\includegraphics[height=0.20 \textheight]{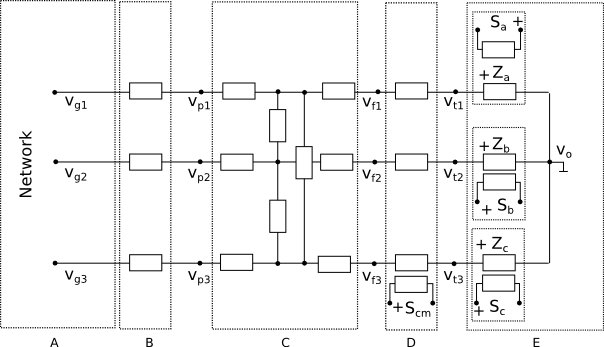} } \quad
	\subfloat[ \label{fig:output_variation}]{\includegraphics[height=0.20 \textheight]{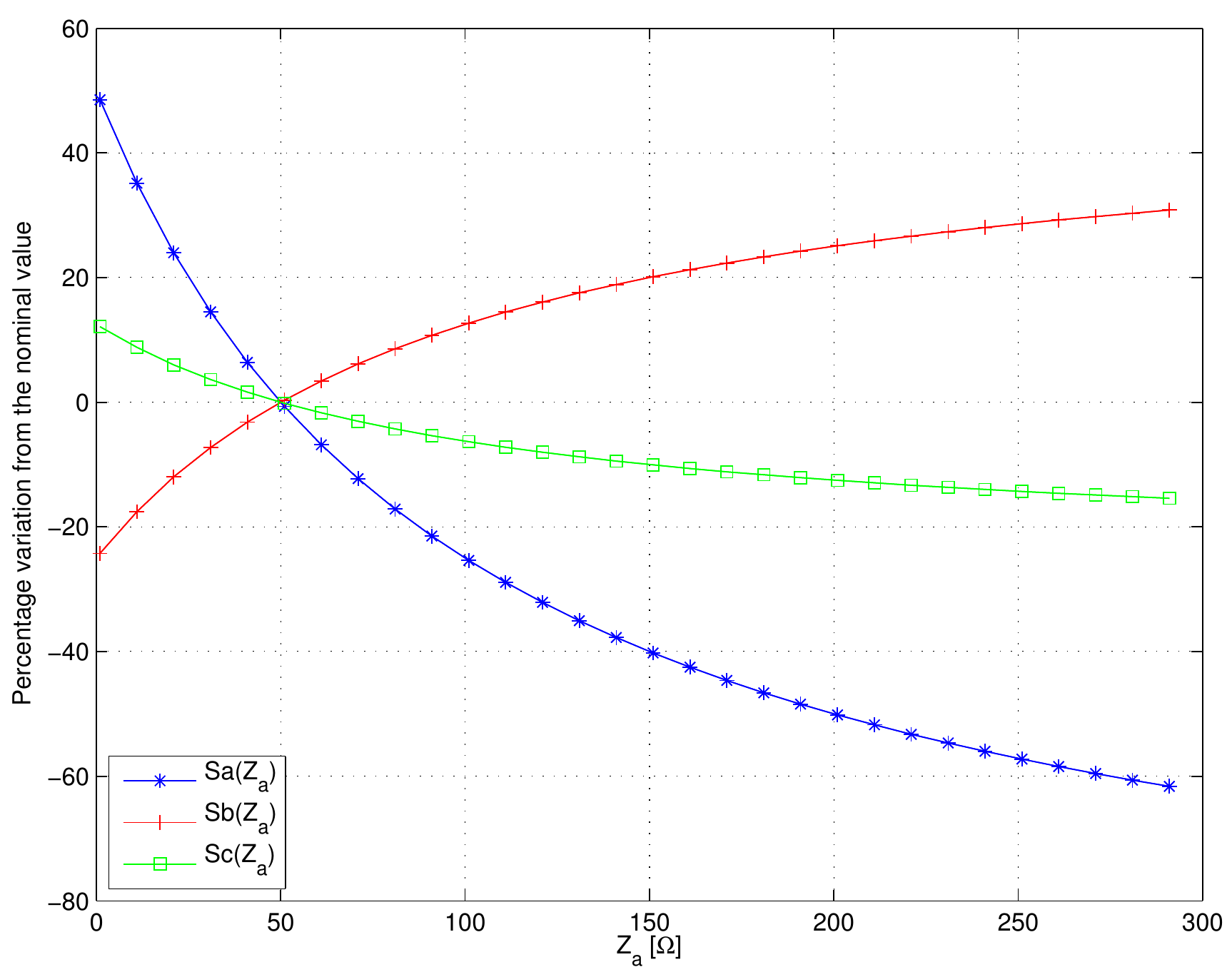} } \\
	\caption{(\ref{fig:star-star_coupler}) Star coupler scheme, (\ref{fig:output_variation}) output variation in percentage from the nominal value of the output signals $S_a, S_b, S_c$.}
	\label{fig:Typ1}
\end{figure}

The behavior of the electrical circuits considered for the MIMO couplers are studied with circuit analysis simulation and analytic models. Each coupler configuration S, $\Delta$, T can be divided into blocks labeled from A to E. As an example, in Fig. \ref{fig:star-star_coupler} a star coupler is shown. Respectively, these blocks represent: the coupler connection to the power line network (A); the over current protection circuitry (B); the high pass filter to attenuate the 50 Hz signal (C); the common mode transformer (D) and uncoupling transformers (E) with loads ($Z_a$,$Z_b$,$Z_c$) that represent the transmitter/receiver impedances.  
The signals that come from the power line network can be modeled with ideal generators with voltages $E_1$,$E_2$,$E_3$ in series with the source impedances $Z_{s1}$,$Z_{s2}$,$Z_{s3}$.
The over current protection circuit is realized with fuses or PTC resettable fuses. The transformers are telecommunications transformers with high bandwidth.
Each of these blocks is described by a transfer function. Exploiting the cascade of the blocks, it is possible to define the overall transfer function of the system.

We have designed and realized a number of couplers for low voltage (LV) networks aiming at analysing the behaviour in the wide band spectrum up to 300 MHz. Some results are reported in the following. 
  
Firstly, we consider, the coupler circuit in Fig. \ref{fig:star-star_coupler}. It comprises the power network inputs and output stages, connected in a star configuration. We can obtain the output voltage signals ($S_a$,$S_b$,$S_c$), considering the input-output stages from A to E as follows:

\small
\begin{equation}
\begin{bmatrix}
i_{a} \\
i_{b} \\
i_{c} \\
\end{bmatrix}
=\frac{1}{D}
\begin{bmatrix}
Z_{s2}+Z_{s3}+Z_{b}+Z_{c} & -(Z_{s3}+Z_{c}) & -(Z_{s2}+Z_{b}) \\
-(Z_{s3}+Z_{a}) & (Z_{s1}+Z_{s3}+Z_{a}+Z_c) & -(Z_{s1}+Z_{a}) \\
-(Z_{s2}+Z_{b}) & -(Z_{s1}+Z_{a}) & (Z_{s1}+Z_{s2}+Z_{a}+Z_b)\\
\end{bmatrix}
\begin{bmatrix}
E_{1} \\
E_{2} \\
E_{3} \\
\end{bmatrix}
\label{eq:star_star_i_basic}
\end{equation}

\begin{equation}
D=(Z_{s1}+Z_{a})(Z_{s2}+Z_{s3}+Z_{b}+Z_{c})+(Z_{s3}+Z_{c})(Z_{s2}+Z_{b}) \notag
\end{equation}

\begin{equation}
\begin{bmatrix}
S_{a} \\
S_{b} \\
S_{c} \\
\end{bmatrix}
=
\begin{bmatrix}
i_{a}Z_{a} \\
i_{b}Z_{b} \\
i_{c}Z_{c} \\
\end{bmatrix}
\label{eq:star_star_v_basic}
\end{equation}
\normalsize

We refer to the case when all the source impedances ($Z_{s1}$,$Z_{s2}$,$Z_{s3}$) are 50$\Omega$ and the load impedances ($Z_a$,$Z_b$,$Z_c$) are matched as the impedance nominal condition. If the source impedances are not perfectly matched to the line impedance, the output signal voltage is severely affected. This is shown in Fig. \ref{fig:output_variation}, where we assume that the source impedances are set to  50$\Omega$ while the load (receiver) impedances are matched with the exception of $Z_a$. The output voltage signals ($S_a$,$S_b$,$S_c$) are calculated as a function of the impedance $Z_a$ variation. The figure shows that a change of $Z_a$ in the range from 0$\Omega$ to 300$\Omega$, cause an output voltages variation up to 60\% from their nominal values.

\begin{figure}[ht!]
	\centering 
	\subfloat[ \label{fig:comparazione_coupler_config}]{\includegraphics[width=0.36 \textheight]{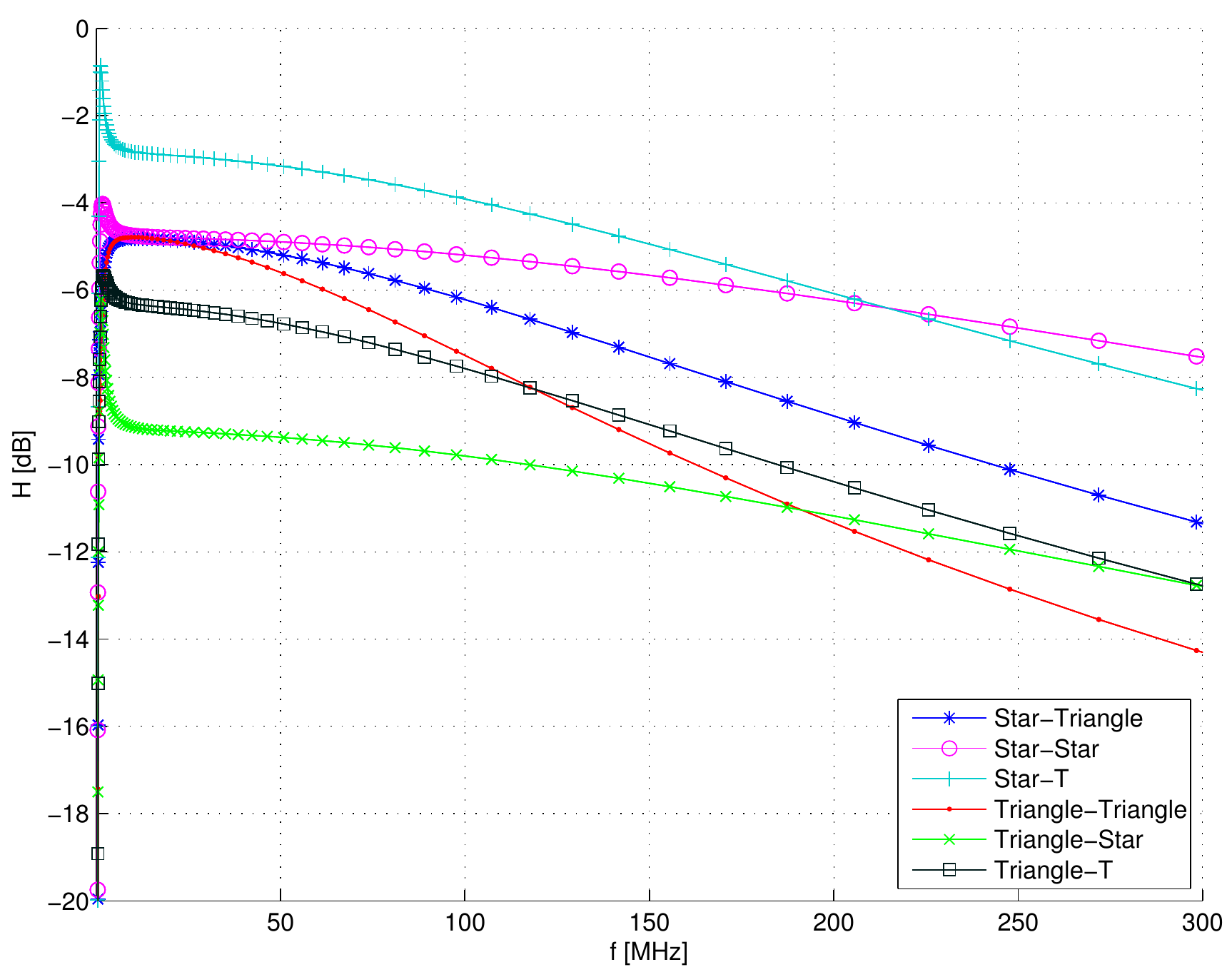} } \quad
	\subfloat[ \label{fig:output_noise_variation_param_zline}]{\includegraphics[width=0.36 \textheight]{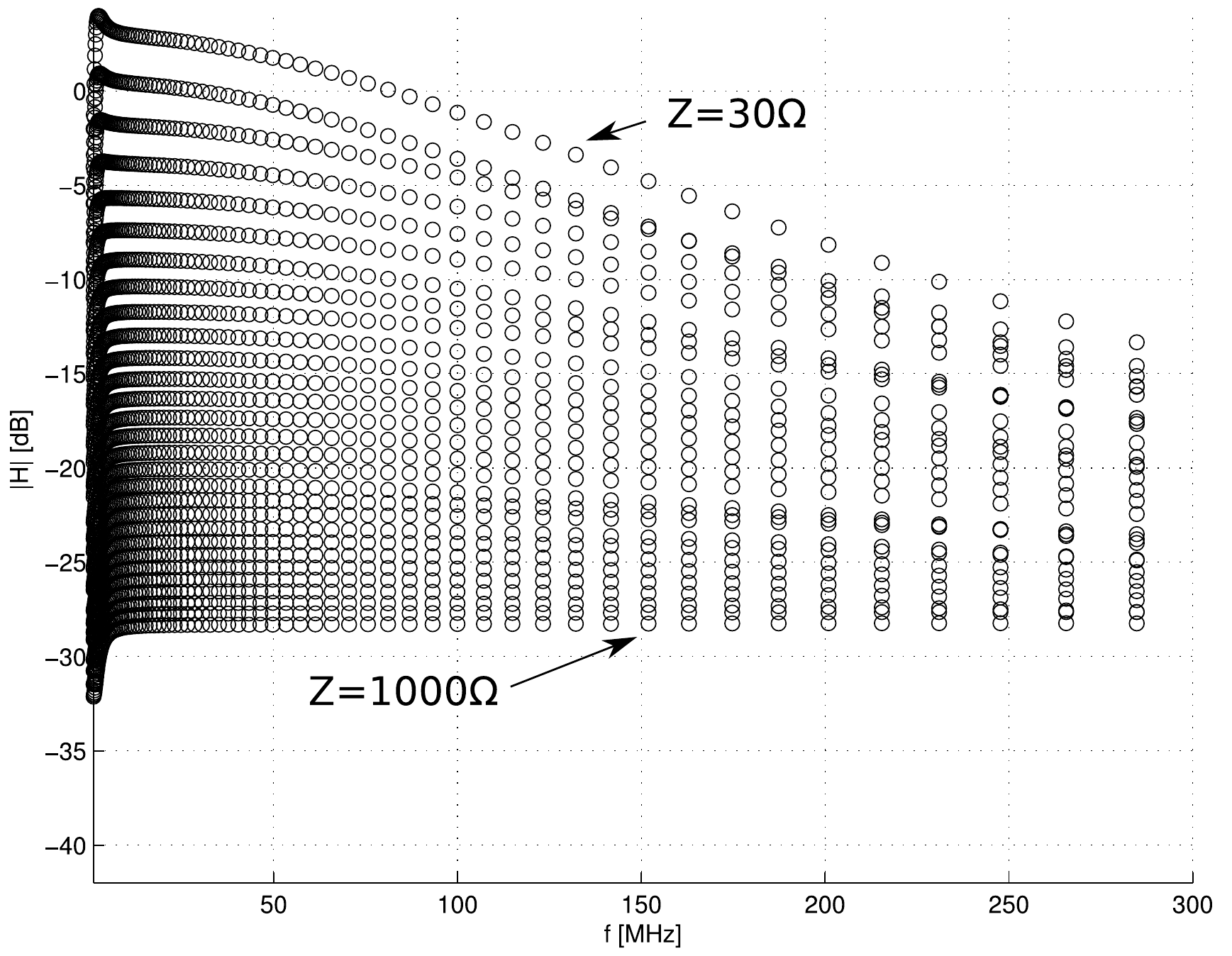} } \\
	\caption{(\ref{fig:comparazione_coupler_config}) Coupler configurations comparison, (\ref{fig:output_noise_variation_param_zline}) transfer functions of triangle-T (transmitter-receiver) coupler configuration, for different values of network impedance.}
	\label{fig:Typ2}
\end{figure}

\begin{figure}[ht!]
\centering 
\includegraphics[width=0.36 \textheight]{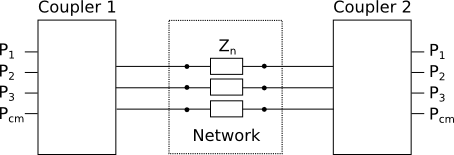}
\caption{Back-to-back coupler connection.}
\label{fig:Typ3}
\end{figure}

Fig. \ref{fig:comparazione_coupler_config}, shows the comparison between the six different back-to-back coupler configurations under analysis (connected as shown in Fig.\ref{fig:Typ3}), considering the  B, C and D sections fixed.
Since, in this analysis we are interested only in the couplers characterization, the power line network is modelled with three network impedances, as shown in Fig.\ref{fig:Typ3}.
The best transfer function characteristic is realized with the transmitter coupler connected in star configuration and the receiver coupler connected in T configuration. However, this coupling can not be practically deployed because of the EMI problems caused by the common mode signal injection. The feasible configurations with the higher transfer functions magnitude, are triangle-T or triangle-triangle.

Fig. \ref{fig:comparazione_coupler_config}, shows the magnitude of the transfer functions with the triangle-T (transmitter-receiver) coupler configuration using different network impedances in the range from 30$\Omega$ to 1000$\Omega$. The highest transfer function magnitude corresponds to the 30$\Omega$ network impedance case, whereas the lowest corresponds to the 1000$\Omega$ case, with a variation of approximately 30dB.
Even more significant variations of the transfer function are expected in real scenarios where the impedance is significantly frequency selective \cite{8}.

Conversely, although not shown, it has been found that the variation of the parameters (passive components of the coupler), from the nominal value, has a negligible effect on the transfer function.

\section{Conclusion}
The analysis here performed shows that the PLC MIMO coupler configuration has a significant impact. Although common mode transmission is not possible for EMI reasons, the common mode signal generated by the network asymmetries and noise, can be exploited at the receiver side.
The combination of a $\Delta$ coupler at the transmitter and $T$ coupler at the receiver has shown to be the best.

\end{document}